\documentclass[aps,prl,twocolumn,groupedaddress]{revtex4}

\usepackage{amsfonts}
\usepackage{amsmath}
\usepackage{amssymb}
\usepackage{graphicx}
\usepackage{times}
\begin{document}
\title{Measuring measurement}
\author{J.S.\ Lundeen$^{1}$}
\email{j.lundeen1@physics.ox.ac.uk}
\author{A.\ Feito$^{2,3}$}
\author{H.\ Coldenstrodt-Ronge$^{1}$}
\author{K.L.\ Pregnell$^{2,3}$}
\author{Ch.\ Silberhorn$^{4}$}
\author{T.C.\ Ralph$^{5}$}
\author{J.\ Eisert$^{2,3,6}$}
\author{M.B.\ Plenio$^{2,3}$}
\author{I.A.\ Walmsley$^{1}$}
\affiliation{$^{1}$Clarendon Laboratory, Oxford University, Parks Road, Oxford, OX1 3PU, UK}
\affiliation{$^{2}$Institute for Mathematical Sciences, Imperial College London, SW7 2PE,
UK }
\affiliation{$^{3}$QOLS, The Blackett Laboratory, Imperial College London, Prince Consort
Rd., SW7 2BW, UK}
\affiliation{$^{4}$Max-Planck Research Group for Optics, Information and Photonics,
Erlangen, Germany}
\affiliation{$^{5}$Department of Physics, University of Queensland, Brisbane, QLD 4072, Australia}
\affiliation{$^{6}$Institute for Physics and Astronomy, University of Potsdam,
14476 Potsdam, Germany}

\date{\today}
\maketitle

\textbf{Measurement connects the world of quantum phenomena to the world of
classical events. It plays both a passive role, observing quantum systems, and
an active one, preparing quantum states and controlling them. Surprisingly --
in the light of the central status of measurement in quantum mechanics --
there is no general recipe for designing a detector that measures a given
observable \cite{Braginsky}. Compounding this, the characterization of
existing detectors is typically based on partial calibrations or elaborate
models. Thus, experimental specification (i.e. tomography) of a detector is of
fundamental and practical importance. Here, we present the realization of
quantum detector tomography [2--4]: we identify the optimal positive-operator-valued measure
describing the detector, with no ancillary assumptions. This result completes
the triad, state [5--11], process [12--17], and detector
tomography, required to fully specify an experiment. We characterize an
avalanche photodiode and a photon number resolving detector capable of
detecting up to eight photons \cite{loopy2}. This creates a new set of tools
for accurately detecting and preparing non-classical light.}

Von Neumann's postulate of the reduction of the quantum state by measurement
is now generally accepted to be a limiting case of a more general theory of
quantum measurement. However, even within this general theory it is not known
how to incorporate the complete chain of apparatus components in a derivation
of the actual measurement: Braginsky wrote, \textquotedblleft the
Schr\"{o}dinger equation cannot tell us the connection between the design of
the measuring device and the nature of the measurement \cite{Braginsky}%
.\textquotedblright\ Measurement is increasingly becoming a driving component
in quantum technologies such as super-resolution metrology \cite{Resch},
Heisenberg-limited sensitivity \cite{Pryde}, and quantum computing
\cite{Knill}. Input states and dynamical processes are accepted as resources
for quantum technologies and therefore the techniques of quantum state
tomography (QST) [5--11] and quantum process tomography (QPT)
[12--17] 
have been
developed to measure them. A distinct omission is that of the experimental
tomography of detectors, which would enable more accurate classification of
measurement types, objective comparison of competing devices, and precise
design of new detectors. This omission is even more striking given that the
tomography of states and processes are predicated on a well characterized
detector. In this paper, we extend previous theoretical descriptions of
detector tomography [2--4] and,
by means of efficient numerics based on convex optimization \cite{Convex} we
characterize two quantum detectors.

Characterizing a detector consists of determining its corresponding
\textit{positive operator valued measure} (POVM). Given an input state $\rho$,
the probability $p_{n,\rho}$ of obtaining detection outcome $n$ is
\begin{equation}
p_{n,\rho}=\mathrm{tr}[\rho\;\pi_{n}], \label{Trace}%
\end{equation}
where $\{\pi_{n}\}$ is the detector POVM. In state tomography, an unknown
$\rho$ is characterized by performing a set of known measurements, each on
many identical copies of the state in order to estimate $p_{n}$. From this
estimate one can invert equation (\ref{Trace}) to find $\rho.$ The
interchangeability of $\rho$ and $\pi_{n}$ in equation (\ref{Trace}) shows
that detector tomography plays a dual role to state tomography. Now, measuring
a set of known probe states $\{\rho\}$ allows us to characterize an unknown
detector, and thus find $\{\pi_{n}\}$. For these operators to describe a
physical measurement apparatus, they must be positive semi-definite, $\pi
_{n}\geq0$, and $\sum_{n}\pi_{n}=I,$ ensuring positive probabilities that add
up to one. In addition, the operators $\{\rho\}$ must be chosen to be
\textit{tomographically complete}, i.e. form a basis for the operator space of
$\pi_{n}$.

In the specific case of optical detectors, lasers provide us with an ideal
tomographic probe: the coherent state $|\alpha\rangle$. By transforming the
magnitude $\left\vert \alpha\right\vert $ through attenuation (e.g. with a
beamsplitter) and the phase $\arg\left(  \alpha\right)  $ by optical delay, we
can create a tomographically complete set of probe states $\{|\alpha
\rangle\langle\alpha|\}$ (the existence of the $P$-function is a proof of
completeness). Remarkably, with coherent state probes, the measured statistics
are themselves a full representation of the detector in the form of the
$Q$-function \cite{PRLSoto},
\begin{equation}
Q_{n}(\alpha)=\frac{1}{\pi}\langle\alpha|\pi_{n}|\alpha\rangle=\frac{1}{\pi
}p_{n,\alpha}. \label{probpure}%
\end{equation}
Since $Q_{n}(\alpha)$ of each POVM element contains the same information as
the element $\pi_{n}$ itself, this is already detector tomography. Predictions
of the detection probabilities for arbitrary input states can then be
calculated directly from the $Q$-function representation. Unfortunately,
experimental errors and statistical fluctuations can cause a simple fit to the
$Q$-function to be consistent with unphysical POVM elements. Due to this we
ultimately wish to directly find the POVM elements $\{\pi_{n}\}$ that are
closest to the measured statistics, while constraining them to be physical.

We now turn to the description of the experimental realization, shown in
Fig.\ 2 (see methods). The first detector was a commercial single-photon
counting module based on a silicon avalanche photodiode (APD). It has two
detection outcomes, either outputting an electronic pulse (1-click) or not
(0-clicks). Past evaluation of the detector has shown that the 1-click outcome
is mainly associated with the arrival of one or more photons, although dark
counts and afterpulsing can also create this outcome. The 0-click event is
mainly associated with vacuum at the input or photons lost due to non-unit
efficiency of the photodiode. Having only two outcomes, this detector cannot
directly measure the incoming photon number if it is above one. The second
detector circumvents this by splitting the incoming pulse into many spatially
or temporally separate bins, making unlikely the presence of more than one
photon per bin. Subsequently all the bins are detected with two APDs.
Photon-number resolution results by summing the number of 1-click outcomes
from all the bins. This \textit{time-multiplexed detector}\ (TMD) is not
commercially available but can be constructed with standard tools
\cite{loopy2}. Ours has eight bins in total (four time bins in each of two
output fibres) and thus nine outcomes -- from zero to eight clicks, making it
capable of detecting up to eight photons. The added complexity and greater
number of outcomes of this detector provide a more challenging test for
detector tomography.

For both detectors we first allowed the phase of $\alpha$ to drift. We
observed no variation in the outcome frequencies, as expected from a detector
without a phase-reference. This simplifies the experimental procedure,
requiring us to control only the magnitude of $\alpha$ (as has been done for
tomography of a single photon \cite{Lvovsky}). A detector with no observed
phase dependence will be described by POVM elements diagonal in the number basis,%

\begin{equation}
\pi_{n}=\sum_{k=0}^{\infty}\theta_{k}^{(n)}|k\rangle\langle k|,
\end{equation}
simplifying henceforth the reconstruction of $\pi_{n}$.

For a POVM set $\{\pi_{n}\}$ containing only diagonal matrices that are each
truncated at a number state $M$, we can rewrite equation (\ref{probpure}) as a
matrix equation,
\begin{equation}
P=F\;\Pi.
\end{equation}
For an $N$ outcome detector, $P_{D\times N}$ contains all the measured
statistics, $F_{D\times M}$ contains the $D$ probe states $\alpha,\alpha
_{1,}\ldots,\alpha_{D}$, and $\Pi_{M\times N}$ contains the unknown POVM set
(matrix subscripts are the matrix dimensions). For a coherent state probe,
$F_{i,k}={|\alpha_{i}|^{2k}\exp{(-|\alpha_{i}|^{2})}}/{k!}$. This can
easily be reformulated for a probe in a mixed state, as was done to model the
laser technical noise (see methods). The optimal physical POVM consistent with
the data can be estimated through the following optimization problem:
\begin{align}
&  \text{min}\left\{  ||P-F\Pi||_{2}+g(\Pi)\right\}  ,\nonumber\label{optim}\\
&  \text{subject to }\;\;\pi_{n}\geq0,\,\;\;\sum_{n=0}^{N-1}\pi_{n}=I,
\end{align}
where the 2-norm of a matrix $A$ is defined as $||A||_{2}=(\sum_{i,j}%
|A_{i,j}|^{2})^{1/2}$. Note that we allow for regularization in the form of
convex quadratic functions $g$, related to the conditioning of the problem,
which must not depend on the type of detector. This is a convex quadratic
optimization problem, and hence also a semi-definite problem (SDP)
\cite{Convex} which can be efficiently solved numerically. Moreover, in this
case, there exists a dual optimization problem whose solution coincides with
the original problem. Thus, the dual problem provides a certificate of
optimality that we use to verify our solution.

\begin{figure}
\includegraphics[width=8.3cm]{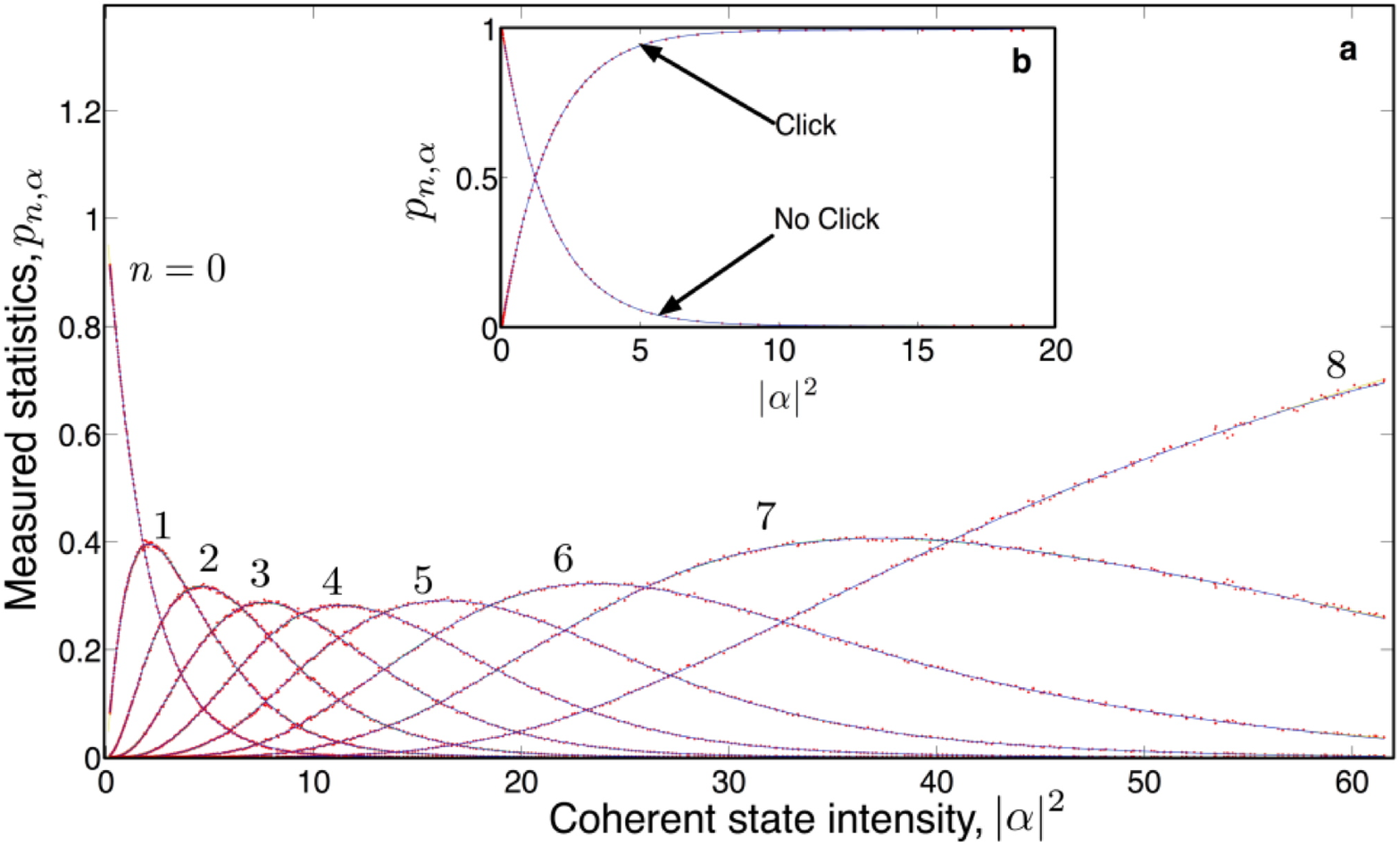}
\caption{The detector tomography data. The outcome statistics (red dots) are
measured as a function of the coherent state magnitude $|\alpha|^{2}$ and form
an estimate of $p_{n,\alpha}$ for each detector outcome $n$ (number of
clicks). Since they are proportional to the $Q$-function $Q_{n}(\alpha)$ for
each outcome, the statistics directly fully characterize the detector. The
main plot corresponds to time multiplexed detector (TMD) with nine outcomes
and the inset corresponds to the avalanche photodiode (APD). The vertical
statistical error is too small to be seen. From the reconstructed POVM
elements $\left\{  \pi_{n}\right\}  $ we generate the corresponding
probabilities $p_{n,\alpha}=\langle\alpha|\pi_{n}|\alpha\rangle$ (blue curves).
}
\end{figure}

\begin{figure}
\includegraphics[width=7cm]{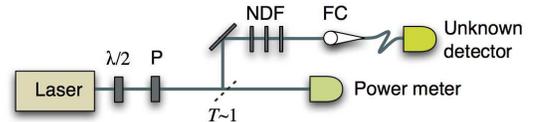}
\caption{The experimental setup. A half-waveplate ($\lambda/2$) and
Glan-Thompson polarizer (P) are used to vary the amplitude of the probe
coherent state, which is subsequently attenuated by Neutral Density Filters
(NDF) and coupled into a fibre (FC) (see methods for more details).
}
\end{figure}

The measured statistics for each detector outcome (i.e. number of clicks) are
shown in Fig.\ 1 for the TMD and for the APD. The distributions (equivalent to
the $Q$-function $Q_{n}(\alpha)$ of the detector) show smooth profiles and
distinct photon number ranges of sensitivity for increasing number of
\textit{clicks} in the detector. Fig.\ 3 shows the diagonals (the
off-diagonals are zero for these phase insensitive detectors) of the POVMs
that result from optimization of equation (\ref{optim}) (see methods for
$g(\Pi)$). Note that $\pi_{n},$ being the POVM element for $n$ clicks, shows
nearly zero amplitude for detecting less than $n$ photons, exhibiting
essentially no dark counts. Prominent in an otherwise smooth distribution,
this sharp feature provides the detector with its discriminatory power: $n$
clicks guarantees there were at least $n$ photons in the input pulse. To
assess the performance of the tomography we find the difference (yellow bars
in Fig. 3) between the estimated POVM elements $\pi_{n}^{\mathrm{rec}}$ and a
previously developed simple theoretical model of a TMD, $\pi_{n}%
^{\mathrm{teo}}$ \cite{AchillesJMO} (see methods). The fidelity 
\begin{equation*}
F=\mathrm{tr}%
\left(  \left(  \sqrt{\pi_{n}^{\mathrm{teo}}}\pi_{n}^{\mathrm{rec}}\sqrt
{\pi_{n}^{\mathrm{teo}}}\right)^{\frac{1}{2}}\right)  ^{2}\geq98.7\% 
\end{equation*}
for all $n$, indicating excellent agreement between the two.

\begin{figure}
\includegraphics[width=8.5cm]{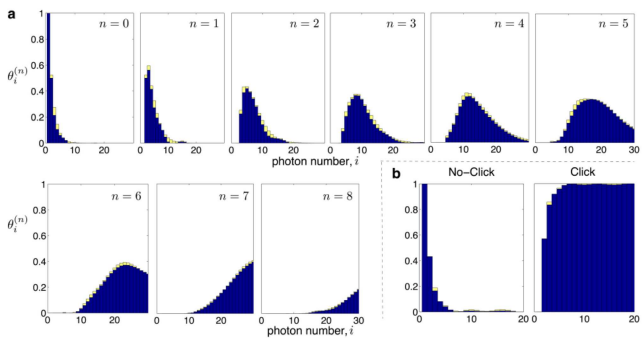}
\caption{The optimal physical POVMs. We present the diagonals of the
reconstructed POVMs represented in the photon-number basis for (a) the
photon-number resolving TMD and (b) the binary APD\ detector. The TMD POVM
elements were obtained up to basis state $|60\rangle\langle60|$ (therefore
$M=60$), but are shown up to $|30\rangle\langle30|$ for display purposes. The
APD POVM elements are shown in full. Stacked on top of each $\theta_{i}^{(n)}$
we show $|\theta_{i}^{(n)\mathrm{(rec)}}-\theta_{i}^{(n)\mathrm{(teo)}}|$ in
yellow, where $n$ is the number of clicks, and rec and teo are the
reconstructed and theoretical diagonals of POVM element, $\pi_{n}$. The
theoretical TMD and APD models are described in methods.
}
\end{figure}

To visualize the action of the detector, in the special case of optical
detectors one can plot a Wigner function of each of the reconstructed POVM
elements, $W_{n}\left(  \alpha,\alpha^{\ast}\right)  $. The response of the
detector to an input state with Wigner function $W_{\psi}$ is proportional to
the overlap, 
\begin{equation*}
p_{n,\psi}=
{\displaystyle\int}
W_{n}W_{\psi}d\alpha d\alpha^{\ast}. 
\end{equation*}
We focus on the one click Wigner
function $W_{1}\left(  \alpha,\alpha^{\ast}\right)  $ for the APD (Fig.\ 4a)
and the TMD (Fig. 4b). An APD detector is sometimes regarded as a `single
photon detector' but here we can see the marked difference between the two
Wigner functions. Instead, it is the TMD that has a fidelity of $98\%$ with a
single photon (having experienced a $52.2\%$ loss). Conversely, the APD Wigner
function extends to $\alpha\gg1,$ having significant overlap with photon
number states $\geqslant1.$ Therefore, to use an APD as a 'single photon
detector' one must make the ancillary assumption that the input beam has
insignificant components containing more than one photon. Despite their
differences, both Wigner functions have negative values near the origin,
indicating the absence of a classical optical analogue. Consequently, these
are both fundamentally quantum detectors.

\begin{figure}
\includegraphics[width=7cm]{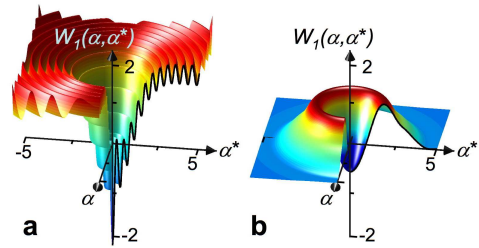}
\caption{The Wigner functions of the `one click' detector outcomes. From the
diagonal elements of $\pi_{1}$ for the APD (a) and TMD (b) one can generate
the Wigner function representing their measurement of the optical mode.
}
\end{figure}

As quantum technologies advance, detectors are becoming more complex, making a
black-box approach to their characterization an important tool. Identifying
the exact operations of detectors will benefit precision tasks, such as state
tomography or metrology. By eliminating assumptions, full characterization
enables more flexible design and use of detectors, be they noisy, nonlinear,
inefficient, or operating outside their normal range. With precise
characterization we can ask precise quantitative questions about our power to
prepare non-classical states or herald quantum operations \cite{Knill}. This
opens a path for the experimental study of yet unexplored concepts such as the
non-classicality of detectors. For optical detector tomography, a promising
avenue for research will be to transfer well-established techniques from
homodyne tomography (e.g. balanced noise-reduction, direct measurement of the
Wigner function or pattern functions \cite{Leonhardt}). Now that it is well
characterized, the photon counter also provides a unique tool for performing
non-Gaussian operations, which are critical for quantum information processing
using the electromagnetic field as the information carrier \cite{Gaussian1,
NonGaussian1}. As superconducting and semiconductor photon number counters are
developed, tomography could be used as an objective benchmark to compare
competing devices. Moreover, for one of these photon number counters only an
incomplete and empirical model is available \cite{Shields}, making detector
tomography the best option to completely determine its action. We expect
detector tomography will become the standard for the adequate calibration of
all measurement and state preparation devices.

\section*{Methods}

\subsection*{Experimental setup}

The pulses of a mode-locked laser travel through a half-waveplate ($\lambda
/2$) and a Glan-Thompson polarizer (P) with which we varied their amplitude
$\alpha$. We subsequently sent the pulses through a beamsplitter (BS)
($T=95\%$). The reflected beam travelled through three neutral density (i.e.
spectrally flat) filters (NDF) before being coupled into a single-mode fiber
(FC). The attenuation from all elements, the reflection off the beamsplitter,
each of the filters, and fibre-coupling, were measured individually with a
calibrated power meter, resulting in a total attenuation $\gamma$. This power
meter was then placed in the transmission port of the beamsplitter so that the
magnitude of $\alpha$ for the probe state in the fibre was found from $P$, the
measured time-averaged power and the pulse rate $R$ via $\left\vert
\alpha\right\vert ^{2}=$ $\gamma P\lambda/(2\pi R\hbar c)$. For each value of
$\alpha$ we recorded the number of times each detection outcome occurred in
$J$ trials (i.e. laser pulses), which provides an estimate of $p_{n,\alpha}$.

\subsection*{Source of light and technical noise}

The input states were generated by a mode-locked Ti:Sapph laser with center
wavelength $\lambda$ and a FWHM bandwidth of $\Delta\lambda$ specifically
chosen for each detector. It was cavity dumped to reduce its repetition rate
$R$ in order to be compatible with tested detectors. Long term drift of the
intensity over 1 million pulses was
%TCIMACRO{\TEXTsymbol{<} }%
%BeginExpansion
$<$
%EndExpansion
0.5\%. To characterize it, a NIST calibrated Coherent FieldMaxII-TO power
meter was used (systematic error of 5\%). In the case of the APD detector (a
Perkin Elmer SPCM-AQR-13-FC) we set $\lambda=780\pm1$ nm, $\Delta\lambda=20$
nm, and chose the appropriate rate $R=1.4975\pm0.0005$ kHz, $J=1472967$, and
$\gamma=(5.66\pm0.08)\times10^{-9}$. For the TMD detector we set
$\lambda=789\pm1$ nm, $\Delta\lambda=26$ nm, $R=76.169\pm0.001$ kHz,
$J=38084$, and $\gamma=(8.51\pm0.11)\times10^{-9}$. We now evaluate the
importance to our tomography of the technical noise found at some level in all
lasers. Our laser randomly varies in energy between subsequent pulses with a
standard deviation of $1.88\%\pm0.02\%$ of $\left\vert \alpha\right\vert ^{2}%
$. Attenuated to the signal photon level, as in this experiment, one might
expect the inherent large fractional uncertainty in the coherent state to
render this technical noise insignificant. We test this expectation by
modelling the pulse distribution as a Gaussian $f_{\alpha}(\beta
)=e^{-(\beta-\alpha)^{2}/(2\sigma^{2})}/(\sigma\sqrt{2\pi})$ centered around
$\alpha$ in phase space, with a variance approximately equal to that measured,
$\sigma^{2}=0.0004\left\vert \alpha\right\vert ^{4}.$ Each probe state is then
best described by a mixture of coherent states,
\begin{align}
\rho_{\left\langle \alpha\right\rangle }  &  =\int d^{2}\beta|\beta
\rangle\langle\beta|f_{\alpha}(\beta)\\
&  =\sum_{l,m=0}^{\infty}E_{l,m,\alpha}|l\rangle\langle m|,
\end{align}
where
\[
E_{j,m,\alpha}=\frac{1}{\sigma\sqrt{2\pi}\sqrt{l!m!}}\int\beta^{l+m}%
e^{-\beta^{2}-(\beta-\alpha)^{2}/(2\sigma^{2})}\;\;d\beta.
\]
The detection probability for outcome $n$ is then
\begin{equation}
p_{\left\langle \alpha\right\rangle ,n}=\sum_{k=0}^{\infty}E_{k,k,\alpha
}\theta_{k}^{(n)}. \label{probmix}%
\end{equation}
Comparing our analysis done with pure input states $|\alpha\rangle
\langle\alpha|$ to that done with mixed states $\rho_{\left\langle
\alpha\right\rangle }$ we find the difference between the POVMs obtained was
negligible. For example
\[
\frac{||\Pi_{\mathrm{pure}}-\Pi_{\mathrm{mixed}}||_{2}}{||\Pi_{\mathrm{mixed}%
}||_{2}}\leq0.7\%
\]
and the largest relative difference between any two $\theta_{k}^{(n)}$ coming
from a mixed state or a pure state derivation was $1.3\%$. Furthermore the
reconstructed probability distributions are so close that they are
indistinguishable on the scale of Fig. 1. This reinforces our earlier
expectation that technical noise in the laser will be negligible when using
single-photon-level coherent states. This differs from homodyne tomography
where technical noise can shift a strong local oscillator to a nearly
orthogonal state.

\subsection*{Discussion of regularization}

Care has to be taken that the optimization problem is well conditioned in
order to find the true POVM of the detector. In finding the number basis
representation we are deconvolving a coherent state from our statistics, which
is intrinsically an ill-conditioned problem. Similar issues of conditioning
have been discussed in the context of state and process tomography, see e.g.
Refs.\ \cite{UnstableBoulant, UnstableHradil}. Due to a large ratio between
the largest and smallest singular values of the matrices defining the
quadratic problem, small fluctuations in the probability distribution can
result in large variations for the reconstructed POVM. This can result in
operators that closely approximate the outcome statistics and yet contain
errant spikes in their distribution in photon-number. To suppress this effect,
we penalize the difference $\theta_{k}^{(n)}-\theta_{k+1}^{(n)}$ (independent
of the shape of the POVM) by using the regularization $g=y\mathcal{S}$ with
$\mathcal{S}=\sum_{k,n}[  \theta_{k}^{(n)}-\theta_{k+1}^{(n)}]
^{2}$. This is motivated by the fact that any realistic detector will have a
finite efficiency $\eta$, which necessitates a smooth $\theta_{k}^{(n)}$
distribution: if $G(r)$ is the probability of registering $r$ photons and
$H(q)$ is the probability that $q$ were present then, 
\begin{equation*}
	G(r)=
{\displaystyle\sum\limits_{q}}
\binom{q}{r}\eta^{r}(1-\eta)^{q-r}H(q).
\end{equation*}
Consequently, if $\theta_{k}\neq0$
then $\theta_{k+1}$, $\theta_{k+2}$ etc. cannot be zero, but will follow some
smooth distribution. Since we do not assume any knowledge about the precise
loss of our detector we simply choose an arbitrary value for $y$. Varying $y$
by three orders of magnitude hardly affects the exact value of the estimated
POVM, changing it by only $10\%$. Furthermore, the regularization
$g=y\;\mathcal{S}$ also proves to be robust to noise up to $\delta=0.2$
(varying $\alpha\rightarrow\alpha(1+\delta)$ across $\{|\alpha\rangle
\langle\alpha|\}$ with a Gaussian distribution for $\delta$). This shows that
the regularization's main effect is to suppress the ill-conditioning and noise
while leaving the POVM fitting unaffected.

\subsection*{ The theoretical model of the detector}

The detector tomography does not make use of any physical model on the
functioning of the detector. To verify the success of this approach we have
compared the outcome of the estimation with those POVM elements obtained from
a theoretical model of the APD and TMD \cite{AchillesJMO}. The APD is treated
as a binary detector with a loss of $43.2\%$. The theoretical TMD assumes: No
dark counts, three sequential beam splitters with experimentally inferred
reflectivities, 50.18\%, 50.60\%, and 41.92\%, and an overall loss of $52.2\%$
(that best fits the data), followed by two perfect APDs.

\section{Acknowledgements}

This work has been supported by the EU integrated project QAP and STREP
COMPAS, the EPSRC grant EP/C546237/1, the EPSRC QIP-IRC, the Royal Society,
Microsoft Research, and the EURYI Award Scheme. HCR has been supported by the
EU under the Marie Curie Prog.\ and by the Heinz-Durr Prog.\ of the
Studienstiftung des dt.\ Volkes.

\section{Competing Financial Interests statement}

The authors have no competing interests as defined by Nature Publishing Group,
or other interests that might be perceived to influence the results and/or
discussion reported in this paper.


\begin{thebibliography}{99}                                                                                               %


\bibitem {Braginsky}V. R. Braginsky and F. Ya. Khalili, \textit{Quantum
measurement}, p. 38 (Cambridge University Press, Cambridge,1992).

\bibitem {PRLSoto}Luis, A. \& Sanchez-Soto, L. L. Complete characterization of
arbitrary quantum measurement processes. Phys. Rev. Lett. \textbf{83,}
3573-3576 (1999).

\bibitem {PhysRevA.64.024102}Fiurasek, J.\ Maximum-likelihood estimation of
quantum measurement. Phys.\ Rev.\ A \textbf{64,} 024102 (2001).

\bibitem {dariano-2004-93}D'Ariano, G.M., Maccone, L. \& Presti, P. L.
Quantum calibration of measurement instrumentation.
Phys.\ Rev.\ Lett.\ \textbf{93,} 250407 (2004).

\bibitem {PhysRevA.40.2847}Vogel, K. \& Risken, H. Determination of
quasiprobability distributions in terms of probability distributions for the
rotated quadrature phase. Phys.\ Rev.\ A \textbf{40,} 2847-2849 (1989).

\bibitem {PhysRevLett.70.1244}Smithey, D. T., Beck, M.,\ Raymer, M. G. \&
Faridani, A. Measurement of the Wigner distribution and the density matrix of
a light mode using optical homodyne tomography: Application to squeezed states
and the vacuum. Phys.\ Rev.\ Lett.\ \textbf{70,} 1244 (1993).

\bibitem {PhysRevA.60.674}Banaszek, K., Radzewicz, C., Wodkiewicz, K. \&
Krasinski, J. S. Direct measurement of the Wigner function by photon counting.
Phys.\ Rev.\ A \textbf{60,} 674 (1999).

\bibitem {Banaszek}Banaszek, K., D'Ariano, G. M., Paris, M. G. A., \& Sacchi,
M. F. Maximum-likelihood estimation of the density matrix. Phys. Rev. A 61,
010304(R) (2000).

\bibitem {White}White, A. G., James, D. F. V., Munro, W. J. \& Kwiat, P. G.
Exploring Hilbert Space: Accurate characterization of quantum information.
Phys. Rev. A \textbf{65,} 012301 (2002).

\bibitem {Grangier}Ourjoumtsev, A., Jeong, H., Tualle-Brouri, R., \& Grangier,
P. Generation of optical `Schr\"{o}dinger cats' from photon number states.
Nature \textbf{448,} 784  (2007).

\bibitem {Polzik}Neergaard-Nielsen, J. S., Melholt Nielsen, B., Hettich, C.,
M\o lmer, K. \& Polzik, E. S. Generation of a superposition of odd photon
number states for quantum information networks. Phys. Rev. Lett. 97, 083604 (2006).

\bibitem {chuang-1996}Chuang, I. L. \& Nielsen, M. A. Prescription for
experimental determination of the dynamics of a quantum black box. J. Mod.
Opt. \textbf{44,} 2455  (1997).

\bibitem {PhysRevLett.78.390}Poyatos, J. F., Cirac, J. I. \& Zoller, P.
Complete characterization of a quantum process: The two-bit quantum gate.
Phys.\ Rev.\ Lett.\ \textbf{78,} 390  (1997).

\bibitem {PhysRevLett.80.5465}D'Ariano, G. M. \& Maccone, L. Measuring quantum
optical Hamiltonians. Phys.\ Rev.\ Lett.\ \textbf{80,} 5465 (1998).

\bibitem {Nielson}Nielsen, M. A., Knill, E., \& Laflamme, R. Complete quantum
teleportation using nuclear magnetic resonance. Nature \textbf{396}, 52  (1998).

\bibitem {kwiat}Altepeter, J. B. et al. Ancilla-assisted quantum process
tomography. Phys.\ Rev.\ Lett.\ \textbf{90,} 193601 (2003).

\bibitem {Mitchell}Mitchell, M. W., Ellenor, C. W., Schneider, S. \&
Steinberg, A. M. Diagnosis, prescription and prognosis of a Bell-state filter
by quantum process tomography. Phys. Rev. Lett. \textbf{91,} 120402 (2003).

\bibitem {loopy2}Achilles, D., Silberhorn, Ch., Sliwa, C., Banaszek, K., \&
Walmsley, I. A. Fiber-assisted detection with photon number resolution. Opt.
Lett. \textbf{28,} 2387  (2003).

\bibitem {Resch}Resch, K. J. et al. Time-reversal and super-resolving phase
measurements. Phys. Rev. Lett. \textbf{98,} 223601 (2007).

\bibitem {Pryde}Higgins, B. L., Berry, D. W., Bartlett, S. D., Wiseman, H. M.
\& Pryde, G. J. Entanglement-free Heisenberg-limited phase estimation. Nature
\textbf{450,} 393  (2007).

\bibitem {Knill}Knill, E. and Laflamme, R. \& Milburn, G. J., A scheme for
efficient quantum computation with linear optics. Nature \textbf{409, }46-52
(2001). 

\bibitem {Convex}Boyd, S. \& Vandenberghe, L. \textit{Convex optimization}
(Cambridge Univ. Press, Cambridge, 2004).

\bibitem {Lvovsky}Lvovsky, A. I. et al. Quantum state reconstruction of the
single-photon Fock state. Phys.\ Rev. \ Lett. \textbf{87},  050402 (2001).

\bibitem {AchillesJMO}Achilles, D. et al. Photon-number-resolving detection
using time-multiplexing. J. Mod. Opt. \textbf{51}, 1499  (2004).

\bibitem {Leonhardt}Leonhardt, U. \textit{Measuring the quantum state of
light} (Cambridge Univ. Press, Cambridge, 1997).

\bibitem {Gaussian1}Eisert, J., Scheel, S., \&\ Plenio, M. B. Distilling
Gaussian states with Gaussian operations is impossible.
Phys.\ Rev.\ Lett.\ \textbf{89},  137903 (2002).

\bibitem {NonGaussian1}Browne, D. E., Eisert, J., Scheel, S. \& Plenio, M. B.
Driving non-Gaussian to Gaussian states with linear optics. Phys. Rev. A
\textbf{67},  062320 (2003).

\bibitem {Shields}Kardynal, B. E., Yuan, Z. L. \& Shields, A. J. An
avalanche-photodiode-based photon-number-resolving detector. Nature Photonics
advance online publication, 15 June 2008 (DOI 10.1038/nphoton.2008.101).

\bibitem {UnstableBoulant}Boulant, N., Havel, T. F., Pravia, M. A. \& Cory, D.
G. Robust method for estimating the Lindblad operators of a dissipative
quantum process from measurements of the density operator at multiple time
points. Phys.\ Rev.\ A \textbf{67}, 042322 (2003).

\bibitem {UnstableHradil}Jezek, M., Fiurasek, J. \& Hradil, Z. Quantum
inference of states and processes. Phys.\ Rev.\ A \textbf{68}, 
012305 (2003).

\end{thebibliography}
\end{document}